%% file: arxiv.tex
\pdfoutput=1
\documentclass[sigconf,nonacm]{acmart}
\usepackage{booktabs}
\usepackage{tikz}
\usetikzlibrary{patterns}
\definecolor{gridblue}{HTML}{8CC4EC}
\definecolor{gridink}{HTML}{0B4F8A}
\input{results}

\setcopyright{none}
\makeatletter\let\hyxmp@parse@acmart\relax\makeatother 
\begin{document}
\title{What a Memory Preview Cannot Tell You: Ambiguity in AI Memory Permissions}
\author{\texorpdfstring{\href{mailto:sadiatonni1916@gmail.com}{Sadia Noman Tonni}}{Sadia Noman Tonni}}
\affiliation{\department{Department of Computer Science and Engineering}\institution{United International University}\city{Dhaka}\country{Bangladesh}}
\author{\texorpdfstring{\href{mailto:nafisfuadshahid@gmail.com}{Nafis Fuad Shahid}}{Nafis Fuad Shahid}}
\affiliation{\department{Department of Computer Science and Engineering}\institution{Islamic University of Technology}\city{Gazipur}\country{Bangladesh}}
\renewcommand{\shortauthors}{Tonni and Shahid}
\begin{abstract}
A memory rule states when an assistant may reuse a remembered event, and an editor can confirm a rule by previewing one use it allows and one it blocks. We ask what such a preview lets a person infer about the rule they saved. For an editor with six or twelve valid rules, we enumerate every rule compatible with a preview. Splitting the rules most evenly does not target the question a user cares about. For an explicit preference, two adaptive tests leave all three recommendation scopes open (for a one-off choice the outcome turns on a tie), while one test of transfer to another task settles the question. Weak previews are also the default. Under the intended rule, 36 of 39 possible allowed/blocked pairs contain a blocked example that no rule allows, and no pair identifies the rule. Three or four tests suffice here; when permissions depend on each other, a fixed checklist grows to six or nine and adaptive questioning needs at most five or seven. These counts use a specified rule vocabulary and fictional events. In a small pilot, 10 of 20 people who saw such a weak preview of a rule they had just chosen mispredicted, or were unsure about, at least one of two uses, and none of those 10 said the outcomes matched what they wanted. Whether better previews help people is untested.
\end{abstract}
\maketitle

\section{Introduction}
A desk bought to fit a temporary workspace is useful context for a matching shelf. Should the purchase also guide a recommendation for a new home? An editor can confirm a memory rule by previewing one use it preserves and one it prevents. Suppose the preview permits the shelf and blocks the claim that the buyer generally prefers pale wood. A rule limited to the workspace and a rule allowing any task both produce those answers, so the preview hides the difference that matters to the buyer.

We ask which future requests distinguish the rules consistent with a preview. For a specified editor we enumerate the compatible rules and report three findings. (1) An even split does not target the user's question: for an explicit preference two such tests leave every recommendation scope open, and for a one-off choice a tie decides, while one test of transfer to another task settles it. (2) Weak previews are the default. Most allowed/blocked pairs an editor could show contain a blocked example that eliminates no rule. (3) Three or four well-chosen tests identify the rule exactly, and the count grows once permissions depend on each other. The enumeration is elementary; the design question is which examples a preview should contain. In a pilot with 20 people, half mispredicted or were unsure about what the rule they had just saved would do.

\begin{figure*}[t]\centering\small
\begin{minipage}[t]{.51\linewidth}\raggedright\textbf{Diagnostic: rules open, next test}\\\footnotesize\emph{Before any answer, for the desk event}\\[3pt]\fbox{\includegraphics[width=\dimexpr\linewidth-2\fboxsep-2\fboxrule\relax]{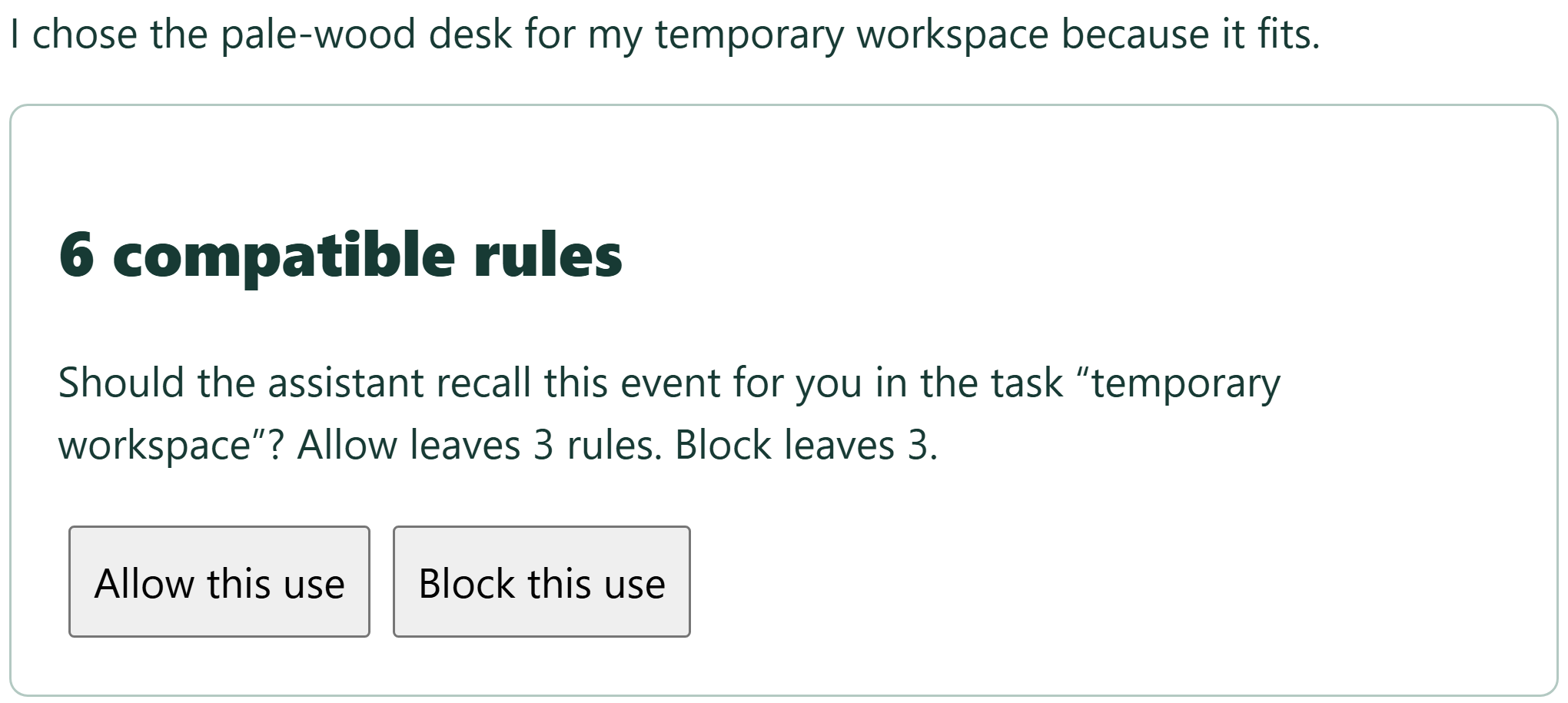}}\end{minipage}\hspace{.03\linewidth}%
\begin{minipage}[t]{.38\linewidth}\raggedright\textbf{Transfer test: use prevented}\\\footnotesize\emph{``What furniture should I buy for my new home?''}\\[3pt]\fbox{\includegraphics[width=\dimexpr\linewidth-2\fboxsep-2\fboxrule\relax]{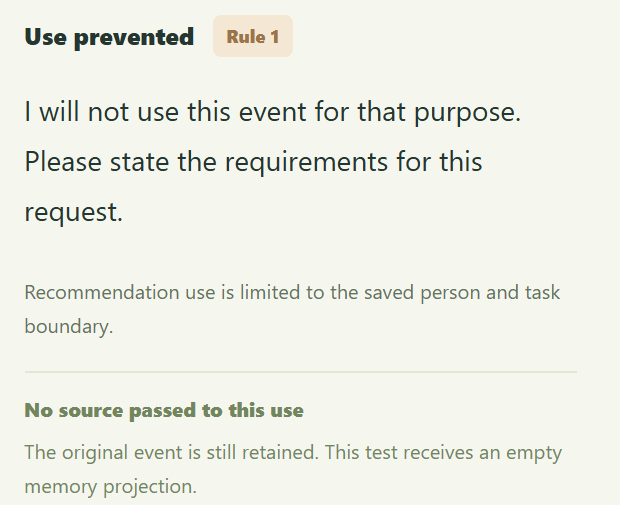}}\end{minipage}
\caption{Left: the diagnostic counts the rules still compatible and suggests a test. Right: the outcome of the transfer test under a saved rule limited to the original task, a request the authored previews did not include. Both are cropped from author-scripted captures.}\label{f:paired}
\Description{Two cropped panels from the prototype. On the left, the diagnostic shows the event text, the heading 6 compatible rules, a suggested question asking whether the assistant should recall the event for you in the temporary workspace task, the note that Allow leaves 3 rules and Block leaves 3, and two buttons, Allow this use and Block this use. On the right, a request about furniture for a new home is marked Use prevented, with a reply that will not use the event and a note that no source was passed.}
\end{figure*}

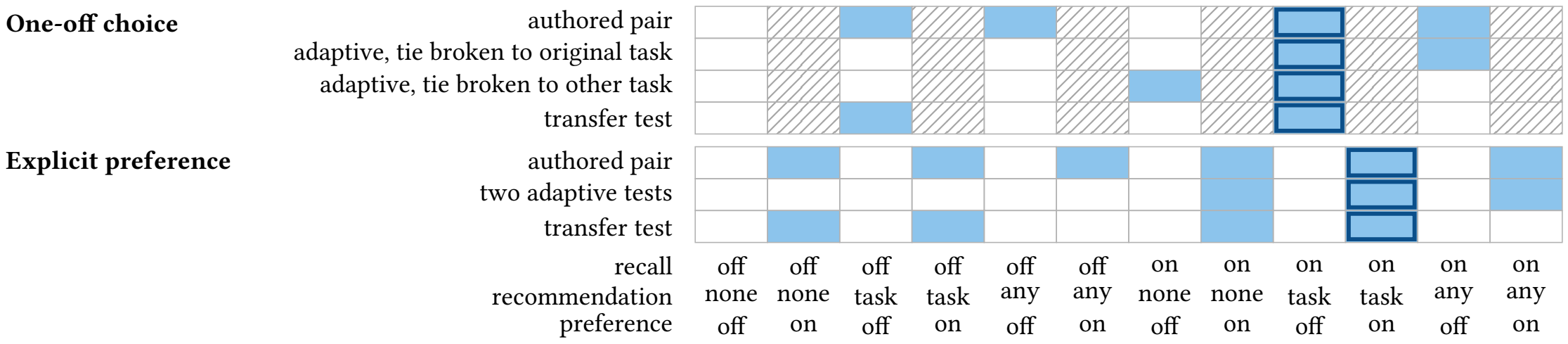
\begin{figure*}[t]\centering
\input{grid-figure}
\caption{Rules still compatible (shaded) after each way of testing; columns are rules, the intended rule is outlined and hatched cells are invalid for a one-off choice. Adaptive tests leave every scope for an explicit preference; for a one-off choice a tie decides.}\label{f:grid}
\Description{A grid with twelve columns, one per rule. For a one-off choice there are four rows. The authored pair leaves four rules. Two adaptive tests with the tie broken to the original task leave two rules that differ only in recommendation scope, task or any. With the tie broken to the other task they leave two rules with scope none or task. The transfer test leaves two rules with scope task that differ only in recall. For an explicit preference there are three rows: the authored pair leaves six rules, two adaptive tests leave three rules with recall on, preference on and scope none, task or any, and the transfer test leaves four rules, none with scope any. The intended rule is outlined and shaded in every row.}
\end{figure*}

\input{related}

\section{Editor and diagnostic}
\textbf{Rules.} An event records a person, a task and an evidence type: a one-off choice or an explicit preference. A rule keeps the event and sets three permissions: \emph{recall} (may the assistant state the fact when asked), \emph{recommendation scope} (none, the original task, or any task for the original person) and \emph{preference use} (may the event stand as a lasting preference). Preference use requires explicit evidence and applies only to the original task. Use for another person and trait inference are always blocked, since a choice or stated preference does not establish a personality trait. This gives six valid rules for a one-off choice and twelve for an explicit preference. A request is a person, a task and a purpose (recall, recommendation, preference or trait), confirmed by the user. Two persons, two tasks and four purposes give sixteen candidate requests.

\textbf{Why preview when there are settings?} A preview checks whether the person's reading of a setting matches what executes. ``Only this task'' does not say whether a bookcase for a new apartment continues the task; a concrete request does. The enumerated rules therefore stand for the readings a person might hold: someone who takes ``only this task'' to cover the apartment expects what the any-task rule does.

\textbf{Prototype and diagnostic.} Replies are deterministic templates, and the gate agreed with a separately coded oracle on 10,000 random inputs. The diagnostic (Figure~\ref{f:paired}) keeps every rule that reproduces the answers so far and suggests the request that splits the rest most evenly (maximum entropy, uniform prior, ties in a fixed order). A separate \emph{transfer} probe changes only the task, the desk buyer's concern.

\section{What previews distinguish}
\textbf{Weak previews are the default.} A preview pairs one request the saved rule allows with one it blocks. Under the intended one-off rule (recall on, recommendations for the original task), three requests are allowed and thirteen blocked, so an editor could show 39 previews. In 36 the blocked example is blocked under every valid rule and eliminates nothing; on average 3.2 of six rules remain, and none identifies the rule. For the intended explicit rule the figures are 44 of 48 previews and 6.3 of twelve rules. The cause is that twelve of the sixteen requests (eleven for explicit events) receive the same answer under every rule. Our own twelve authored scenarios fell into this: each prevented example was one that no rule allows, leaving four of six or six of twelve rules (Figure~\ref{f:grid}).

\begin{table}[ht]\caption{Two tests aimed at each intended rule: rules left $K$, residual bits $H$ ($\log_2 K$; a mean for Random) and disagreement $D$.}\label{t:metrics}\centering
\begin{tabular}{lrrr@{\qquad}rrr}\toprule &\multicolumn{3}{c}{One-off choice}&\multicolumn{3}{c}{Explicit preference}\\
&$K$&$H$&$D$&$K$&$H$&$D$\\\midrule\previewrows\\\bottomrule\end{tabular}\end{table}

\textbf{An even split does not target the user's question.} Table~\ref{t:metrics} compares four ways of choosing two tests. \emph{Random} averages all 120 unordered pairs of requests; restricted to distinct requests that can separate rules (requests with identical outcomes under every rule counted once), it leaves 2.0 and 3.8 rules. \emph{Adaptive} takes the diagnostic's suggestion twice. \emph{Transfer} keeps the authored preserved use and adds an original-person recommendation for another task. $D$ is the share of remaining rule pairs that disagree, averaged over the sixteen requests. For an explicit preference the adaptive tests are recall and original-task preference, with no tie; three rules remain, one per recommendation scope (Figure~\ref{f:grid}), so transfer is entirely open, while the transfer test leaves four rules, none allowing another task. A one-off choice depends on a tie: after recall, the original-task and other-task recommendation tests split the rest equally. Our fixed order takes the first and leaves the task-limited and any-task rules; the other answers the buyer's question, with the same $K$, $H$ and $D$.

\textbf{How many tests identify a rule?} Two yes-or-no answers have at most four patterns, so two tests cannot distinguish all six rules in the worst case. Three can: a recall request, an original-task recommendation and an other-task recommendation, all for the original person. Adding an original-task preference request identifies all twelve explicit rules. Both match the lower bound $\lceil\log_2 K\rceil$. That is one test per independent setting. Dependent permissions change this. An offline sensitivity model makes recommendation, handoff and follow-up require recall and adds expiry (never or from day seven), preference scope (none, original or any task; explicit evidence only), a same-task handoff to one designated person and same-task follow-up; a rule permitting nothing has no expiry. This gives 25 and 77 rules. The 48 requests combine day zero or seven, two persons, two tasks and six purposes (recall, recommendation, preference, handoff, follow-up and trait). The smallest \emph{fixed checklist}, chosen in advance to tell every two rules apart, has six and nine requests, above the bounds of five and seven; adaptive questioning needs at most five and seven.

\section{Where a task ends}
Can software decide whether a new request continues the task? We wrote 48 fictional request pairs (24 same task, 24 different; a new beneficiary, goal or occasion means a different task), with provisional labels drafted with AI assistance and reproduced by a second AI pass, which shows internal consistency only. On 24 held-out pairs, literal equality recognized 3 of 12 same-task pairs, a fitted text-embedding-3-small threshold accepted 23 of 24 pairs, and a label-blind Gemini 3.8 Flash judge recognized 11 of 12 with no false positive. The prototype therefore uses task identifiers the person confirms, at a cost: it blocked all 12 archived requests that renamed the task.

\section{A pilot with people}
Twenty adults, classmates and friends of the authors invited in person, completed an on-screen survey on the desk event. Each chose a recommendation scope (six none, nine the original task, five any task), saw a two-example preview whose blocked example no rule allows, and predicted two uses of their rule before seeing the outcomes: a lamp for the workspace and a bookcase for a new apartment. Ten of 20 mispredicted or were unsure about at least one. Five mispredicted the original task; six did not predict the transfer (three wrong, three unsure, the only unsure answers given). Errors ran both ways: four people expected a use their rule blocks and four expected a block their rule allows. None of the ten answered Yes to whether the outcomes matched what they wanted, against nine of the other ten. The sample is small and known to the authors, the median completion time was 41 seconds, and there was no comparison condition, so we report counts only. Participants were unpaid adult volunteers who consented before participation and could stop or discard their responses. No identifying or sensitive information was collected. Formal institutional ethics review was not obtained.

\section{Discussion and next study}
For preview design: draw examples that some rule allows and another blocks, since an always-blocked example checks only the implementation; let the person's stated concern choose a test, since an even split can leave it open; and count the rules a preview leaves open and offer the next test, which here costs three or four tests in total.

The counts describe one finite rule set, and the diagnostic treats every rule as equally likely. Replies are templates, so nothing here shows that a language model would respect a rule. The pilot shows that readings diverge from what executes; it does not show that a better preview closes the gap. The next study should randomly show the authored or the transfer preview and compare predictions, giving people time to read.

\textit{Code and data.}\hfill\break\url{https://github.com/tonni19/memory-preview-ambiguity}

\textit{AI use.} Generative AI tools (OpenAI Codex and Anthropic Claude) were used to write the code, author the fictional fixtures and provisional labels, run the archived model calls, and draft and edit this text. The authors reviewed the manuscript and are responsible for its content.
\bibliographystyle{ACM-Reference-Format}
\bibliography{references}
\end{document}

%% file: results.tex
\newcommand{\previewrows}{%
Authored & 4 & 2.0 & 12.5\% & 6 & 2.6 & 14.2\% \\
Random & 4.8 & 2.2 & 12.3\% & 9.0 & 3.1 & 14.1\% \\
Adaptive & 2 & 1.0 & 6.3\% & 3 & 1.6 & 8.3\% \\
Transfer & 2 & 1.0 & 12.5\% & 4 & 2.0 & 12.5\% %
}

%% file: grid-figure.tex
\begin{tikzpicture}[x=1cm,y=1cm,font=\footnotesize]
\node[anchor=west,font=\footnotesize\bfseries] at (0,-0.150) {One-off choice};
\node[anchor=east] at (6.500,-0.150) {authored pair};
\draw[black!30,line width=.3pt] (6.600,0.000) rectangle (7.280,-0.300);
\path[pattern=north east lines,pattern color=black!35] (7.280,0.000) rectangle (7.960,-0.300);
\draw[black!30,line width=.3pt] (7.280,0.000) rectangle (7.960,-0.300);
\fill[gridblue] (7.960,0.000) rectangle (8.640,-0.300);
\draw[black!30,line width=.3pt] (7.960,0.000) rectangle (8.640,-0.300);
\path[pattern=north east lines,pattern color=black!35] (8.640,0.000) rectangle (9.320,-0.300);
\draw[black!30,line width=.3pt] (8.640,0.000) rectangle (9.320,-0.300);
\fill[gridblue] (9.320,0.000) rectangle (10.000,-0.300);
\draw[black!30,line width=.3pt] (9.320,0.000) rectangle (10.000,-0.300);
\path[pattern=north east lines,pattern color=black!35] (10.000,0.000) rectangle (10.680,-0.300);
\draw[black!30,line width=.3pt] (10.000,0.000) rectangle (10.680,-0.300);
\draw[black!30,line width=.3pt] (10.680,0.000) rectangle (11.360,-0.300);
\path[pattern=north east lines,pattern color=black!35] (11.360,0.000) rectangle (12.040,-0.300);
\draw[black!30,line width=.3pt] (11.360,0.000) rectangle (12.040,-0.300);
\fill[gridblue] (12.040,0.000) rectangle (12.720,-0.300);
\draw[black!30,line width=.3pt] (12.040,0.000) rectangle (12.720,-0.300);
\path[pattern=north east lines,pattern color=black!35] (12.720,0.000) rectangle (13.400,-0.300);
\draw[black!30,line width=.3pt] (12.720,0.000) rectangle (13.400,-0.300);
\fill[gridblue] (13.400,0.000) rectangle (14.080,-0.300);
\draw[black!30,line width=.3pt] (13.400,0.000) rectangle (14.080,-0.300);
\path[pattern=north east lines,pattern color=black!35] (14.080,0.000) rectangle (14.760,-0.300);
\draw[black!30,line width=.3pt] (14.080,0.000) rectangle (14.760,-0.300);
\draw[gridink,line width=1.1pt] (12.070,-0.030) rectangle (12.690,-0.270);
\node[anchor=east] at (6.500,-0.450) {adaptive, tie broken to original task};
\draw[black!30,line width=.3pt] (6.600,-0.300) rectangle (7.280,-0.600);
\path[pattern=north east lines,pattern color=black!35] (7.280,-0.300) rectangle (7.960,-0.600);
\draw[black!30,line width=.3pt] (7.280,-0.300) rectangle (7.960,-0.600);
\draw[black!30,line width=.3pt] (7.960,-0.300) rectangle (8.640,-0.600);
\path[pattern=north east lines,pattern color=black!35] (8.640,-0.300) rectangle (9.320,-0.600);
\draw[black!30,line width=.3pt] (8.640,-0.300) rectangle (9.320,-0.600);
\draw[black!30,line width=.3pt] (9.320,-0.300) rectangle (10.000,-0.600);
\path[pattern=north east lines,pattern color=black!35] (10.000,-0.300) rectangle (10.680,-0.600);
\draw[black!30,line width=.3pt] (10.000,-0.300) rectangle (10.680,-0.600);
\draw[black!30,line width=.3pt] (10.680,-0.300) rectangle (11.360,-0.600);
\path[pattern=north east lines,pattern color=black!35] (11.360,-0.300) rectangle (12.040,-0.600);
\draw[black!30,line width=.3pt] (11.360,-0.300) rectangle (12.040,-0.600);
\fill[gridblue] (12.040,-0.300) rectangle (12.720,-0.600);
\draw[black!30,line width=.3pt] (12.040,-0.300) rectangle (12.720,-0.600);
\path[pattern=north east lines,pattern color=black!35] (12.720,-0.300) rectangle (13.400,-0.600);
\draw[black!30,line width=.3pt] (12.720,-0.300) rectangle (13.400,-0.600);
\fill[gridblue] (13.400,-0.300) rectangle (14.080,-0.600);
\draw[black!30,line width=.3pt] (13.400,-0.300) rectangle (14.080,-0.600);
\path[pattern=north east lines,pattern color=black!35] (14.080,-0.300) rectangle (14.760,-0.600);
\draw[black!30,line width=.3pt] (14.080,-0.300) rectangle (14.760,-0.600);
\draw[gridink,line width=1.1pt] (12.070,-0.330) rectangle (12.690,-0.570);
\node[anchor=east] at (6.500,-0.750) {adaptive, tie broken to other task};
\draw[black!30,line width=.3pt] (6.600,-0.600) rectangle (7.280,-0.900);
\path[pattern=north east lines,pattern color=black!35] (7.280,-0.600) rectangle (7.960,-0.900);
\draw[black!30,line width=.3pt] (7.280,-0.600) rectangle (7.960,-0.900);
\draw[black!30,line width=.3pt] (7.960,-0.600) rectangle (8.640,-0.900);
\path[pattern=north east lines,pattern color=black!35] (8.640,-0.600) rectangle (9.320,-0.900);
\draw[black!30,line width=.3pt] (8.640,-0.600) rectangle (9.320,-0.900);
\draw[black!30,line width=.3pt] (9.320,-0.600) rectangle (10.000,-0.900);
\path[pattern=north east lines,pattern color=black!35] (10.000,-0.600) rectangle (10.680,-0.900);
\draw[black!30,line width=.3pt] (10.000,-0.600) rectangle (10.680,-0.900);
\fill[gridblue] (10.680,-0.600) rectangle (11.360,-0.900);
\draw[black!30,line width=.3pt] (10.680,-0.600) rectangle (11.360,-0.900);
\path[pattern=north east lines,pattern color=black!35] (11.360,-0.600) rectangle (12.040,-0.900);
\draw[black!30,line width=.3pt] (11.360,-0.600) rectangle (12.040,-0.900);
\fill[gridblue] (12.040,-0.600) rectangle (12.720,-0.900);
\draw[black!30,line width=.3pt] (12.040,-0.600) rectangle (12.720,-0.900);
\path[pattern=north east lines,pattern color=black!35] (12.720,-0.600) rectangle (13.400,-0.900);
\draw[black!30,line width=.3pt] (12.720,-0.600) rectangle (13.400,-0.900);
\draw[black!30,line width=.3pt] (13.400,-0.600) rectangle (14.080,-0.900);
\path[pattern=north east lines,pattern color=black!35] (14.080,-0.600) rectangle (14.760,-0.900);
\draw[black!30,line width=.3pt] (14.080,-0.600) rectangle (14.760,-0.900);
\draw[gridink,line width=1.1pt] (12.070,-0.630) rectangle (12.690,-0.870);
\node[anchor=east] at (6.500,-1.050) {transfer test};
\draw[black!30,line width=.3pt] (6.600,-0.900) rectangle (7.280,-1.200);
\path[pattern=north east lines,pattern color=black!35] (7.280,-0.900) rectangle (7.960,-1.200);
\draw[black!30,line width=.3pt] (7.280,-0.900) rectangle (7.960,-1.200);
\fill[gridblue] (7.960,-0.900) rectangle (8.640,-1.200);
\draw[black!30,line width=.3pt] (7.960,-0.900) rectangle (8.640,-1.200);
\path[pattern=north east lines,pattern color=black!35] (8.640,-0.900) rectangle (9.320,-1.200);
\draw[black!30,line width=.3pt] (8.640,-0.900) rectangle (9.320,-1.200);
\draw[black!30,line width=.3pt] (9.320,-0.900) rectangle (10.000,-1.200);
\path[pattern=north east lines,pattern color=black!35] (10.000,-0.900) rectangle (10.680,-1.200);
\draw[black!30,line width=.3pt] (10.000,-0.900) rectangle (10.680,-1.200);
\draw[black!30,line width=.3pt] (10.680,-0.900) rectangle (11.360,-1.200);
\path[pattern=north east lines,pattern color=black!35] (11.360,-0.900) rectangle (12.040,-1.200);
\draw[black!30,line width=.3pt] (11.360,-0.900) rectangle (12.040,-1.200);
\fill[gridblue] (12.040,-0.900) rectangle (12.720,-1.200);
\draw[black!30,line width=.3pt] (12.040,-0.900) rectangle (12.720,-1.200);
\path[pattern=north east lines,pattern color=black!35] (12.720,-0.900) rectangle (13.400,-1.200);
\draw[black!30,line width=.3pt] (12.720,-0.900) rectangle (13.400,-1.200);
\draw[black!30,line width=.3pt] (13.400,-0.900) rectangle (14.080,-1.200);
\path[pattern=north east lines,pattern color=black!35] (14.080,-0.900) rectangle (14.760,-1.200);
\draw[black!30,line width=.3pt] (14.080,-0.900) rectangle (14.760,-1.200);
\draw[gridink,line width=1.1pt] (12.070,-0.930) rectangle (12.690,-1.170);
\node[anchor=west,font=\footnotesize\bfseries] at (0,-1.470) {Explicit preference};
\node[anchor=east] at (6.500,-1.470) {authored pair};
\draw[black!30,line width=.3pt] (6.600,-1.320) rectangle (7.280,-1.620);
\fill[gridblue] (7.280,-1.320) rectangle (7.960,-1.620);
\draw[black!30,line width=.3pt] (7.280,-1.320) rectangle (7.960,-1.620);
\draw[black!30,line width=.3pt] (7.960,-1.320) rectangle (8.640,-1.620);
\fill[gridblue] (8.640,-1.320) rectangle (9.320,-1.620);
\draw[black!30,line width=.3pt] (8.640,-1.320) rectangle (9.320,-1.620);
\draw[black!30,line width=.3pt] (9.320,-1.320) rectangle (10.000,-1.620);
\fill[gridblue] (10.000,-1.320) rectangle (10.680,-1.620);
\draw[black!30,line width=.3pt] (10.000,-1.320) rectangle (10.680,-1.620);
\draw[black!30,line width=.3pt] (10.680,-1.320) rectangle (11.360,-1.620);
\fill[gridblue] (11.360,-1.320) rectangle (12.040,-1.620);
\draw[black!30,line width=.3pt] (11.360,-1.320) rectangle (12.040,-1.620);
\draw[black!30,line width=.3pt] (12.040,-1.320) rectangle (12.720,-1.620);
\fill[gridblue] (12.720,-1.320) rectangle (13.400,-1.620);
\draw[black!30,line width=.3pt] (12.720,-1.320) rectangle (13.400,-1.620);
\draw[black!30,line width=.3pt] (13.400,-1.320) rectangle (14.080,-1.620);
\fill[gridblue] (14.080,-1.320) rectangle (14.760,-1.620);
\draw[black!30,line width=.3pt] (14.080,-1.320) rectangle (14.760,-1.620);
\draw[gridink,line width=1.1pt] (12.750,-1.350) rectangle (13.370,-1.590);
\node[anchor=east] at (6.500,-1.770) {two adaptive tests};
\draw[black!30,line width=.3pt] (6.600,-1.620) rectangle (7.280,-1.920);
\draw[black!30,line width=.3pt] (7.280,-1.620) rectangle (7.960,-1.920);
\draw[black!30,line width=.3pt] (7.960,-1.620) rectangle (8.640,-1.920);
\draw[black!30,line width=.3pt] (8.640,-1.620) rectangle (9.320,-1.920);
\draw[black!30,line width=.3pt] (9.320,-1.620) rectangle (10.000,-1.920);
\draw[black!30,line width=.3pt] (10.000,-1.620) rectangle (10.680,-1.920);
\draw[black!30,line width=.3pt] (10.680,-1.620) rectangle (11.360,-1.920);
\fill[gridblue] (11.360,-1.620) rectangle (12.040,-1.920);
\draw[black!30,line width=.3pt] (11.360,-1.620) rectangle (12.040,-1.920);
\draw[black!30,line width=.3pt] (12.040,-1.620) rectangle (12.720,-1.920);
\fill[gridblue] (12.720,-1.620) rectangle (13.400,-1.920);
\draw[black!30,line width=.3pt] (12.720,-1.620) rectangle (13.400,-1.920);
\draw[black!30,line width=.3pt] (13.400,-1.620) rectangle (14.080,-1.920);
\fill[gridblue] (14.080,-1.620) rectangle (14.760,-1.920);
\draw[black!30,line width=.3pt] (14.080,-1.620) rectangle (14.760,-1.920);
\draw[gridink,line width=1.1pt] (12.750,-1.650) rectangle (13.370,-1.890);
\node[anchor=east] at (6.500,-2.070) {transfer test};
\draw[black!30,line width=.3pt] (6.600,-1.920) rectangle (7.280,-2.220);
\fill[gridblue] (7.280,-1.920) rectangle (7.960,-2.220);
\draw[black!30,line width=.3pt] (7.280,-1.920) rectangle (7.960,-2.220);
\draw[black!30,line width=.3pt] (7.960,-1.920) rectangle (8.640,-2.220);
\fill[gridblue] (8.640,-1.920) rectangle (9.320,-2.220);
\draw[black!30,line width=.3pt] (8.640,-1.920) rectangle (9.320,-2.220);
\draw[black!30,line width=.3pt] (9.320,-1.920) rectangle (10.000,-2.220);
\draw[black!30,line width=.3pt] (10.000,-1.920) rectangle (10.680,-2.220);
\draw[black!30,line width=.3pt] (10.680,-1.920) rectangle (11.360,-2.220);
\fill[gridblue] (11.360,-1.920) rectangle (12.040,-2.220);
\draw[black!30,line width=.3pt] (11.360,-1.920) rectangle (12.040,-2.220);
\draw[black!30,line width=.3pt] (12.040,-1.920) rectangle (12.720,-2.220);
\fill[gridblue] (12.720,-1.920) rectangle (13.400,-2.220);
\draw[black!30,line width=.3pt] (12.720,-1.920) rectangle (13.400,-2.220);
\draw[black!30,line width=.3pt] (13.400,-1.920) rectangle (14.080,-2.220);
\draw[black!30,line width=.3pt] (14.080,-1.920) rectangle (14.760,-2.220);
\draw[gridink,line width=1.1pt] (12.750,-1.950) rectangle (13.370,-2.190);
\node[anchor=east] at (6.500,-2.440) {recall};
\node at (6.940,-2.440) {off};
\node at (7.620,-2.440) {off};
\node at (8.300,-2.440) {off};
\node at (8.980,-2.440) {off};
\node at (9.660,-2.440) {off};
\node at (10.340,-2.440) {off};
\node at (11.020,-2.440) {on};
\node at (11.700,-2.440) {on};
\node at (12.380,-2.440) {on};
\node at (13.060,-2.440) {on};
\node at (13.740,-2.440) {on};
\node at (14.420,-2.440) {on};
\node[anchor=east] at (6.500,-2.720) {recommendation};
\node at (6.940,-2.720) {none};
\node at (7.620,-2.720) {none};
\node at (8.300,-2.720) {task};
\node at (8.980,-2.720) {task};
\node at (9.660,-2.720) {any};
\node at (10.340,-2.720) {any};
\node at (11.020,-2.720) {none};
\node at (11.700,-2.720) {none};
\node at (12.380,-2.720) {task};
\node at (13.060,-2.720) {task};
\node at (13.740,-2.720) {any};
\node at (14.420,-2.720) {any};
\node[anchor=east] at (6.500,-3.000) {preference};
\node at (6.940,-3.000) {off};
\node at (7.620,-3.000) {on};
\node at (8.300,-3.000) {off};
\node at (8.980,-3.000) {on};
\node at (9.660,-3.000) {off};
\node at (10.340,-3.000) {on};
\node at (11.020,-3.000) {off};
\node at (11.700,-3.000) {on};
\node at (12.380,-3.000) {off};
\node at (13.060,-3.000) {on};
\node at (13.740,-3.000) {off};
\node at (14.420,-3.000) {on};
\end{tikzpicture}

%% file: related.tex
\section{Related work}
A version space is the set of hypotheses consistent with the examples seen~\cite{mitchell}, and the teaching dimension is the fewest examples that single one out~\cite{goldman}. Program synthesis queries distinguishing inputs~\cite{jha}, programming-by-example interfaces elicit such distinctions from users~\cite{mayer,peleg,regae}, Expandable Grids exposes effective permissions~\cite{grids}, and policy-by-example learns rules from cases~\cite{n}. We ask which contrasts a memory-rule preview must contain.

For assistant memory, user studies describe memory that departs from people's expectations~\cite{chen} and disclosures that travel beyond their intended context~\cite{taheri}; contextual integrity frames such reuse as a flow that must fit its context~\cite{barth}, and benchmarks measure violations~\cite{ci,persist}. Memory Sandbox lets users view and edit an agent's memories~\cite{sandbox}, other systems govern what is admitted~\cite{gated,liu} or filter records by provenance and permission~\cite{mapgraph}, and access-control models grant by attributes~\cite{rbac,abac}. The closest work is ConMem~\cite{conmem}, which synthesizes counterfactual scenarios so that a system can audit whether to retain a memory. We use counterfactual requests for a different reader: the person checking a rule they wrote. Yan~\cite{yan} found that user-authored standing policies blocked less agent overreach than per-action review, because most rules deferred to ``ask''. A working policy editor is not evidence of benefit.